\documentclass[preprint,aps,nofootinbib]{revtex4}

\usepackage{graphicx}
\newcommand{\beq}{\begin{equation}}
\newcommand{\eeq}{\end{equation}}
\newcommand{\bqa}{\begin{eqnarray}}
\newcommand{\eqa}{\end{eqnarray}}

\begin{document}


\title{Implications for hydrodynamic fluctuations on the  minimum shear viscosity of the dilute Fermi gas at unitarity}

\author{Paul Romatschke and Ryan Edward Young}

\affiliation{Department of Physics, 390 UCB, University of Colorado, Boulder,
CO 80309, USA}

\begin{abstract}
We confirm and expand on 
work by Chafin and Sch\"afer on hydrodynamic fluctuations
in the unitary Fermi gas. Using the result for the equation
of state from a recent MIT experiment, we derive lower bounds
for $\eta/n$ and $\eta/s$ as a function of temperature.
Re-analyzing recent quantum Monte Carlo data for the shear-viscosity 
spectral function we point out a possible resolution for the tension 
between the viscosity bound $\eta/n\gtrsim 0.3$ from Chafin and Sch\"afer and the quantum Monte Carlo results $\eta/n\lesssim 0.2$ from Wlazlowski et al.
near the critical temperature.
\end{abstract}

\maketitle

\section{Motivation}

Hydrodynamics allows for long lived shear and sound waves.  Fluctuations can populate these modes and the interactions of these waves can dissipate momentum.  This contributes to the shear viscosity and other transport coefficients. 

Our study builds upon results found in the context of relativistic fluids,
specifically Ref.~\cite{Kovtun:2011np}, which was recently applied to the case of the Fermi gas at unitarity by Chafin and Sch\"afer in Ref.~\cite{Chafin:2012fu}. In Ref.~\cite{Chafin:2012fu} it was found that the dimensionless combination of shear viscosity over density $\eta/n$ must be larger than or equal to $0.32$ for the unitary Fermi gas at the superfluid-normal phase transition temperature $T=T_c$.
Thus, there is a tension between this result and the result for $\eta/n$ found in the quantum Monte Carlo simulations by Wlazlowski et al. in Ref.~\cite{Wlazlowski:2012jb}, who found $\left.\eta/n\right|_{T\sim T_c}\lesssim 0.2$. This provides the motivation for us to re-derive and generalize the result by Chafin and Sch\"afer as well as re-analyze the results by Wlazlowski et al.

This work is organized as follows. In section \ref{sec:classical}, we give a 
brief review of the differences between the classical and fluctuating (stochastic) 
formulation of hydrodynamics. In section \ref{sec:min}, using results from 
fluctuating hydrodynamics, we derive a result for the minimum physical viscosity 
of a cold quantum gas. In section \ref{sec:width}, we re-analyze quantum Monte 
Carlo simulation data from Ref.~\cite{Wlazlowski:2012jb} and compare 
the resulting spectral functions to our analytic low-frequency results. 
We conclude in section \ref{sec:discussion} and discuss our use of the equation of state from a recent MIT experiment in Appendix A. 
Note that throughout our work we use natural units $\hbar=k_B=c=1$, unless stated otherwise.

\section{Classical versus Fluctuating Hydrodynamics}
\label{sec:classical}

In the following, we will distinguish between quantities calculated in ``classical'' hydrodynamics (bare quantities) and quantities calculated in fluctuating hydrodynamics (physical quantities). Since some readers may not be familiar with these concepts, we will briefly review what we mean by these terms.

Classical hydrodynamics is the low frequency/low wavenumber effective theory of energy momentum conservation. It can be derived from first principles by first identifying the relevant degrees of freedom (fluid velocity, particle density) and the central object (the energy momentum tensor). One then proceeds to write down all possible terms in the energy momentum tensor allowed by symmetry, ordered by the number of gradients involved. The most compact way of achieving this is using Poincar\'e symmetry and relativistic 4-vector notation (see e.g. Ref.~\cite{Romatschke:2009im} for a review), but all the building blocks are present in classic fluid dynamics textbooks, such as Ref.~\cite{LL}. In the non-relativistic context, to zeroth order in the gradient expansion, the energy density is given by the fluid mass density, $T_{00}=\rho$, the momentum density is given by mass density times fluid velocity, $T_{0i}=\rho v_i$, and the stress tensor is given by $T_{ij}=\rho v_i v_j + P \delta_{ij}$ with $P$ the fluid pressure. Energy and momentum conservation then read
\begin{eqnarray}
\partial_t T_{00}+\partial_j T_{0j}&=&\partial_t \rho + \partial_j \left(\rho v_j\right)=0\,\nonumber\\
\partial_t T_{0i}+\partial_j T_{ij}&=&\partial_t  \left(\rho v_i\right) + \partial_j \left(\rho v_i v_j + P \delta_{ij}\right)=0\,,
\end{eqnarray}
which, upon some rearrangement, can be seen to represent the equation of continuity and Euler equations, respectively. Inclusion of derivative terms up to first order in gradients in the energy-momentum tensor leads to
$$
T_{ij}=\rho v_i v_j + P \delta_{ij} - \eta_{\rm cl} \left(\partial_i v_j+\partial_j v_i-\frac{2}{3}\delta_{ij}\partial_k v_k\right) - \zeta_{\rm cl} \delta_{ij} \partial_k v_k\,,
$$
where $\eta_{\rm cl},\zeta_{\rm cl}$ are the classical shear and bulk viscosity coefficients, respectively. It is straightforward to verify that the corresponding conservation equations are the Navier-Stokes equations. Higher-order gradient term corrections can be calculated, making the approach systematic, but these will be of no relevance to the present work and hence will not be discussed here.

It is well known that this classical hydrodynamic description, though very successful in its own right, does not do justice to the dynamics of real fluids because it does not do full justice to the fluctuation-dissipation theorem (see e.g. Ref.~\cite{Kovtun:2012rj} for a recent review). However, it is well known how to lift this shortcoming, namely by the introduction of a stochastic noise term in the energy-momentum tensor:
$$
T_{ij}\rightarrow T_{ij}=T_{ij}^{\rm cl}+\Xi_{ij}\,,\quad
\langle \Xi_{ij}(x)\Xi_{lm}(y)\rangle \propto \delta^4(x-y)\,,
$$
(see appendix \ref{sec:finding} for an illustration), where here and in the following the notation ``cl'' will denote classical quantities. This formulation of ``fluctuating'' hydrodynamics contains equations of motion of classical hydrodynamics as the average of the equations of motion over the noise, e.g.
\begin{equation}
\langle \partial_t T_{0i}+\partial_j T_{ij}\rangle =  \partial_t T_{oi}^{\rm cl}+\partial_j T_{ij}^{\rm cl}\,,
\end{equation}
because the nature of the noise correlator. However, the two formulations will differ when calculating correlation functions, such as 
the retarded correlator of the energy-momentum tensor component $T^{xy}$, namely
$$
G^{R}_{xyxy}(x-y)=\langle T_{xy}(x) T_{xy}(y)\rangle_R\,.
$$
This correlator is of special interest because its Fourier transform can be related to the value of transport coefficients. For simplicity, let us restrict our treatment to the special case of a conformal system with a vanishing bulk viscosity coefficient. Then, in classical hydrodynamics, the Fourier transform of the above mentioned correlator becomes (\cite{Chafin:2012fu}, cf.~\cite{Baier:2007ix})
\beq
\label{eq:correlator}
G^{R,{\rm cl}}_{xyxy}(\omega,{\bf k})=P-i \eta_{\rm cl} \omega + \eta \tau^{\rm cl}_{R} \omega^2-\frac{\kappa^{\rm cl}}{2}{\bf k}^2 + {\cal O}(\omega^3,{\bf k}^3)\,,
\eeq
with $\eta_{\rm cl}$ the classical shear viscosity coefficient as before. Here $\tau_R^{\rm cl}$ (the relaxation time) and $\kappa^{\rm cl}$ are two transport coefficients that arise when considering hydrodynamics at second order in fluctuations (one order higher than Navier-Stokes). They will turn out to be irrelevant for the following discussion but have been included here for completeness.

As mentioned above, the stochastic formulation of hydrodynamics contains this result as a special case (with the exception of the frequency and momentum independent ``contact'' term $P$ in $G^{R,{\rm cl}}_{xyxy}$). This is most easily seen by first realizing that at zero momentum, $G_{xyxy}^R$ is related to another energy-momentum tensor correlator via an hydrodynamic analogue of the Ward-Takahashi identity\footnote{This can be shown easily by considering correlators built out of $\partial_t T_{i0}=-\partial_j T_{ij}$. }
$$
G^{R}_{xyxy}(\omega,0)=\lim_{{\bf k}\rightarrow 0}\frac{\omega^2}{{\bf k}^2}\langle T_{0y}T_{0y}\rangle_R\,,
$$
if we take ${\bf k}=(k,0,0)$. Using $T_{0y}=\rho v_t$, a short calculation given in appendix \ref{sec:finding} leads to the finding that
\emph{to lowest order in fluctuations} 
$$
\langle T_{0y}T_{0y}\rangle_R^{(0)}=\rho_o^2 \langle v^t_y v^t_y\rangle_R
=\frac{-\eta_{\rm cl} k^2}{-i \omega+\gamma k^2}
$$
which in turn leads to 
\begin{equation}
G^{R,(0)}_{xyxy}(\omega,0)=-i \omega \eta_{\rm cl}\,.
\end{equation}
Note that this result matches the classical hydrodynamic result Eq.~(\ref{eq:correlator}) up to terms of first order in frequency. This is because it was derived using the hydrodynamic formulation including only first order gradients in the energy-momentum tensor (commonly known as Navier-Stokes equations). If higher order versions of hydrodynamics are employed, the terms involving $\tau_R,\kappa$ in Eq.~(\ref{eq:correlator}) should also be described correctly. 

Moreover, note that this result was derived keeping only the lowest order contribution in fluctuations. Clearly, correction to this result involving higher order contributions from fluctuations will also contribute to the full result. Hence
$$
G^{R}_{xyxy}(\omega,0)=G^{R,(0)}_{xyxy}(\omega,0)+G^{R,(1)}_{xyxy}(\omega,0)+G^{R,(2)}_{xyxy}(\omega,0)+\ldots\,,
$$
where as above $G^{R,(0)}=G^{R,{\rm cl}}$. Therefore, correlation functions in fluctuating hydrodynamics will in general differ from those calculated in classical hydrodynamics. 

Let us evaluate the first correction term to the classical result in the above series to illustrate the result. A short calculation shows that at this order, the correlator receives a contribution in the form of a momentum integral over two hydrodynamic propagators $\Delta_{ij}=\langle v_i v_j \rangle$, e.g. it is a one-loop integral in the language of field theory. Without repeating the detailed calculation that can be found in  Ref.~\cite{Kovtun:2011np}, we only state the result, which turns out to be
\begin{eqnarray}
G^{R,(1)}_{xyxy}(\omega, 0) & = &\rho^2 \int \frac{ d \omega'}{2 \pi} \int \frac{d^3 \mathbf{p}}{(2 \pi)^3} \times \nonumber\\ && \bigg( \Delta_S^{xx}(\omega', \mathbf{p}) \Delta_R^{yy}(\omega-\omega', -\mathbf{p})+  \Delta_S^{xy}(\omega', \mathbf{p})\Delta_R^{yx}(\omega-\omega', -\mathbf{p})+\nonumber \\ && \Delta_S^{xx}(\omega', \mathbf{p}) \Delta_R^{yy}(\omega-\omega', -\mathbf{p})+\Delta_S^{xy}(\omega', \mathbf{p})\Delta_R^{yx}(\omega-\omega', -\mathbf{p})  \bigg)\,,
\end{eqnarray}
where $\Delta^{ij}_{S,R}$ are the symmetric and retarded version of the velocity-velocity correlator (see appendix \ref{sec:finding}). In the zero momentum limit $|{\bf k}|\rightarrow 0$, the integration of the one-loop contribution to the correlator is straightforward (cf.~\cite{Kovtun:2011np}), and one finds
\begin{equation}
\label{eq:oneloop}
G^{R,(1)}_{xyxy}(\omega, 0) (\omega \ll p_{\rm max},|{\bf k}|=0) ={\rm const} -i\omega \frac{17T p_{max}}{120\gamma \pi^2}+(1+i) \omega^{3/2} \frac{(28+3 \sqrt{6})T}{980 \pi \gamma^{3/2}}+\dots\,,
\end{equation}
where $\gamma=\frac{\eta}{\rho}$ and $p_{\rm max}$ is the cut-off for the effective theory of hydrodynamics. The leading constant term is a renormalization of the pressure and does not affect transport, hence we will ignore it in the following. Note that terms not included in Eq.~(\ref{eq:oneloop}) are those that are suppressed by higher powers of $p_{\rm max}\gamma$, which is the ``small parameter'' of the effective theory. Similarly, higher loop contributions to $G_R^{xyxy}$ are suppressed by higher powers of $p_{\rm max}\gamma$. Thus, up to first order in powers of $p_{\rm max}\gamma$, the result for the retarded correlator is given by the sum of the ``classical'' result Eq.~(\ref{eq:correlator}) and the one-loop result Eq.~(\ref{eq:oneloop}).

\section{Minimum on viscosity}
\label{sec:min}

Using the Kubo relation 
\begin{equation}
\label{Kubo}
\eta=\lim_{\omega \rightarrow 0}\frac{- {\rm Im}G^R_{xyxy}(\omega,{\bf k}=0)}{\omega}\,,
\end{equation}
the result for the physical viscosity including fluctuations to first loop order is given by
\begin{equation}
\label{etafull}
\eta=\eta_{\rm cl}+\frac{17 T p_{\rm max} \rho}{120 \eta_{\rm cl} \pi^2}\,,
\end{equation}
where again the index ``cl'' refers to ``classical'' contributions in the sense of turning off all fluctuations. As was previously noted, our calculation is systematic if
\begin{equation}
\label{criterion}
p_{\rm max} \ll \frac{\rho}{\eta_{\rm cl}}\,,
\end{equation}
so we parametrize $p_{\rm max}$ by
$$
p_{\rm max}=\frac{\rho}{\eta_{\rm cl}} \Phi\,,
$$
where $\Phi$ is a dimensionless function of thermodynamic variables that 
should be small. For instance, we may take\footnote{In principle, $\Phi$ may also depend on the combination $T/T_F$ with $T_F$ the Fermi temperature (see below). However, we will drop this dependence in favor of a constant that may vary around unity.} $\Phi$ to be a function of pressure over density, $\Phi=\Phi(P/\rho)$. 
Then the result for the physical viscosity may be cast into a result for the dimensionless viscosity over density ratio,
$$
\left(\frac{\eta}{n}\right)=\left(\frac{\eta}{n}\right)_{\rm cl}
+\frac{17 T \rho^2 \Phi(P/\rho)}{120 n^3 \left(\eta/n\right)^2_{\rm cl} \pi^2}\,.
$$
This expression can be extremized. We find that the physical viscosity over density expression may never decrease below the minimum value
\beq
\label{viscmin}
\left(\frac{\eta}{n}\right)\geq \left(\frac{\eta}{n}\right)_{\rm min}=
\left(\frac{153 T \rho^2 \Phi(P/\rho)}{160 \pi^2 n^3}\right)^{1/3}\,.
\eeq
The minimum value for viscosity is dimensionless and presumably universal, meaning
that for a unitary Fermi gas it should only depend on the combination $T/T_F$ with $T_F=\frac{(3\pi^2 n)^{2/3}}{2 m}$. Assuming a power-like behavior, we have
$$
\left(\frac{153 T \rho^2 \Phi(P/\rho)}{160 \pi^2 n^3}\right)\propto\left(\frac{T}{T_F}\right)^\alpha\,,
$$
with $\alpha$ a constant. Similarly, we take $\Phi$ to be given by a power law with behavior
$$
\Phi(P/\rho)\propto \left(P/\rho\right)^\beta\,,
$$
with $\beta$ a constant. As shown in appendix \ref{sec:eos}, the pressure is 
given by the relation $P= n T G(T/T_F)$ and $\lim_{T\rightarrow \infty} G=1$. With this, we find
$$
\frac{T m^2}{n} \left(\frac{T}{m}\right)^\beta \propto \left(\frac{T}{T_F}\right)^\alpha\,,
$$
which leads to the requirement $\alpha=1+\beta$\,, $2-\beta=\alpha$ and
thus
$$
\alpha=\frac{3}{2}\,,\quad \beta=\frac{1}{2}\,.
$$
Thus, we obtain the final form of the lower viscosity bound
\beq
\label{viscmin2}
\left(\frac{\eta}{n}\right)\geq \left(\frac{\eta}{n}\right)_{\rm min}=
\left(k \frac{153 T \rho^2  \sqrt{P/\rho})}{160 \pi^2 n^3}\right)^{1/3}\,,
\eeq
with $k$ an arbitrary constant of order unity (see the discussion in section \ref{sec:3b}). One finds
\begin{equation}
\label{eq:viscmin3}
\left(\frac{\eta}{n}\right)\geq \left(\frac{\eta}{n}\right)_{\rm min}=
\left(k \frac{459\sqrt{\frac{T^3}{T_F^3} \frac{P}{nT}}}{2^{3/2}160  }\right)^{1/3}\simeq 1.005 k^{1/3} \sqrt{\frac{T}{T_F}} \left(\frac{P}{n T}\right)^{1/6}\,,
\end{equation}
which matches the result found in Ref.~\cite{Chafin:2012fu}.

\subsection{Calculating the classical viscosity}

Given a liquid with a measurable energy-momentum tensor component $T^{xy}$,
the shear viscosity of that liquid can be defined using the 
Kubo relation.
Since this (physical) viscosity will in general differ from the classical viscosity $\eta_{\rm cl}$ we have defined above, this may cause some confusion. Thus we will take this as an opportunity to clarify how the two quantities will differ and point out an example of how to obtain both of these quantities.

Let us start by pointing out that the real, physical viscosity $\eta$ of a system is what one obtains if one performs a measurement of the correlator $G^R_{xyxy}$ from experiment and then uses Eq.~(\ref{Kubo}) to obtain $\eta$. Similarly, if one was able to simulate a fully quantum non-equilibrium system and unambiguously extract the correlator from this simulation, the Kubo formula of this quantity would yield the physical shear viscosity. Since at the time of writing this is prohibited by the so-called sign problem, let us investigate what happens if one turns to the usual kinetic theory method to extract the shear viscosity. For this purpose, consider a scalar quantum field with a quartic self-interaction with coupling constant $\lambda$. It is well known that in the limit of weak coupling $\lambda\rightarrow 0$, the dynamics of this quantum field is well approximated by kinetic theory (cf. Ref.~\cite{Blaizot:2001nr}):
\begin{equation}
\label{Boltzmann}
\partial_t f({\bf p},{\bf x},t) + \frac{1}{E} {\bf p \cdot \nabla} f({\bf p},{\bf x},t) = - 
{\cal C}[f]\,,
\end{equation}
where $f({\bf p},{\bf x},t)$ is the phase-space distribution of particles with momentum ${\bf p}$ and energy $E$, and ${\cal C}[f]$ is the collision kernel that is proportional to the transport cross section, ${\cal C}[f]\sim \lambda^2$. It is also well known that the description in terms of kinetic theory breaks down once the (quasi-)particle states start to have a width that is comparable to their mass, which typically happens when the coupling $\lambda$ starts to become sizable.

Nevertheless, as long as $\lambda\ll 1$ the description in terms of kinetic theory is applicable, and one can use known techniques such as the Chapman-Enskog expansion or Grad's 14 moment method to extract the shear viscosity $\eta_{KT}$ from the kinetic description. Normalizing by particle density one finds
$$
\frac{\eta_{KT}}{n}\sim \frac{1}{\lambda^2}\,,
$$
up to logarithmic corrections in $\lambda$, and where we have suppressed dimensionful parameters that depend on the problem of interest (mass, temperature, etc.). 

Now consider the contribution to the physical viscosity arising from hydrodynamic fluctuations, Eq.~(\ref{etafull}). If the coupling is weak, we expect kinetic theory to give a good description of the physical viscosity. Sound modes should be short-lived and hence the contribution coming from fluctuating hydrodynamics should be small. Hence, we are led to identify
$$
\eta_{\rm cl}\simeq \eta_{KT}+{\cal O}(\lambda^{-1})\,.
$$
But if this is the case, one may ask the question of why the kinetic theory description misses the contribution to viscosity from fluctuations as the coupling strength $\lambda$ is increased. Plugging in the above estimates one finds
$$
\eta-\eta_{\rm cl}=\frac{17 T \rho^2 }{120 \eta^2_{\rm cl} \pi^2} \Phi
\propto
\lambda^4 T m^2 \Phi\,.
$$
In other words, the fluctuating contribution to $\eta$ is suppressed by six powers of the coupling strength $\lambda$ with respect to $\eta_{KT}$. Clearly, the original kinetic description Eq.~(\ref{Boltzmann}) with ${\cal C}[f]\propto \lambda^2$ is not accurate to this order in the coupling. Moreover, it is known that to capture effects at this order in the coupling constant, it would not even be sufficient to calculate the cross section or collision kernel ${\cal C}[f]$ correctly up to $\lambda^6$ because the kinetic theory description contains new structures that cannot be brought in the simple form of a Boltzmann equation, cf.~\cite{Weinstock:2005jw}.

To conclude, at weak coupling, a good approximation to the ``classical'' viscosity can be calculated microscopically by using a kinetic theory description. However, care must be taken in so far as this identification relies on properly calculating the weak coupling cross section used in the kinetic theory treatment. Furthermore, the kinetic theory description will become inaccurate as the coupling strength is increased, since the kinetic theory description cannot be systematically improved beyond the leading-order. Comparing the parametric dependence on the coupling $\lambda$, it is likely that $\eta_{KT}$ will become an inaccurate approximation of $\eta_{\rm cl}$ before the physical viscosity is dominated by the contribution arising from fluctuations, e.g. at intermediate values of the coupling strength.

\subsection{Is the expansion systematic?}
\label{sec:3b}

One may worry if the expansion based on Eq.~(\ref{criterion}) is systematic because at first glance both $p_{\rm max}$ as well as $\rho/\eta_{\rm cl}$ are expected to scale as the inverse mean free path. In the case at hand, however, we found that $p_{\rm max}$ is equal to $\rho/\eta_{\rm cl}$ times a dimensionless function $\Phi=\sqrt{P/\rho}$. Our setup is thus systematic as long as
$$
\Phi\sim \frac{1}{v}\sqrt{\frac{k_B T}{m}} \sim \frac{c_{T}}{c}\ll 1\,,
$$
where  $c_{T}$ is the thermal speed of the atoms.
Note that we have restored standard units and included a characteristic system-dependent speed $v$ that is needed to make $\Phi$ dimensionless, as required.
In a relativistic system, $v$ would simply be the speed of light. Since the speed of light does not arise in our non-relativistic treatment, we have to construct another characteristic speed that must not depend on the temperature or on the particle masses (otherwise the resulting function $\Phi$ would not satisfy the scaling requirement outlined in the beginning of this section). We are thus forced to construct a characteristic speed out of the typical system size $L\sim 100 \mu m$  and the period $t=2\pi/\omega$ from the trapping frequency  $\omega\simeq 2\pi\times 2000$ Hz (cf. Ref.~\cite{Adams:2012th}), finding $v\simeq 0.5$ m/s. The typical temperatures considered are on order the Fermi temperature
$$
T_F\simeq \frac{\hbar}{k_B} \omega (6 N)^{1/3}\,,
$$
where the number of atoms is typically $N\simeq 7.5\times 10^4$, hence $T_F\simeq 7.9 \mu K$ and thus for a cold quantum gas of ${\rm Li}^6$ atoms $c_T\simeq 0.1$ m/s. Hence we find
$$
\Phi=\frac{c_T}{v}\simeq 0.2\,,
$$
which is sufficiently small to suggest that higher order corrections are suppressed, but not so different from unity as to put our earlier estimate for the constant $k\simeq 1$ in Eq.~(\ref{eq:viscmin3}) into question. To conclude, for temperatures on the order of $T_F$ there seems to be a small parameter allowing the expansion in powers of fluctuations to be systematic, but we do expect sizable corrections to our leading order result. Therefore, we expect our results based on a leading order analysis to be qualitative robust, but not quantitatively accurate.

\section{Width of Spectral Function Peak}
\label{sec:width}

Collecting all terms of the retarded correlator $G_R^{xyxy}(\omega)$
from Eq. (\ref{eq:correlator}) and Eq. (\ref{eq:oneloop}),
the full result for the one-loop spectral function $\eta(\omega)$ to
leading order in frequency is given by
\begin{eqnarray}
\label{specfunc}
\frac{\eta(\omega)}{n}&=&\left(\frac{\eta}{n}\right)_{\rm cl}+
k \frac{51}{2^{3/2} 120} \left(\frac{\eta}{n}\right)_{\rm cl}^{-2}
\left(\frac{T}{T_F}\right)^{3/2}\sqrt{\frac{P}{n T}}\nonumber\\
&&-\frac{3 \pi(28+3 \sqrt{6})}{2^{3/2} 980}  \left(\frac{\eta}{n}\right)_{\rm cl}^{-3/2} \left(\frac{T}{T_F}\right) \sqrt{\frac{\omega}{T_F}}+{\cal O}(\omega^2)\,,
\end{eqnarray}
with $k$ an unknown constant of order unity.
Let us now assume we have arrived at the viscosity coefficient $\lim_{\omega\rightarrow 0}\frac{\eta(\omega)}{n}$ by some other means, e.g. a first principle lattice calculation.

\noindent The viscosity coefficient then fixes the value of $\eta_{\rm cl}/n$ and hence 
Eq.~(\ref{specfunc}) gives the low-frequency behavior of the spectral function. We may then compare this result to the spectral function obtained by the first-principles method.

\begin{figure}
\begin{center}
\includegraphics[width=0.45\linewidth]{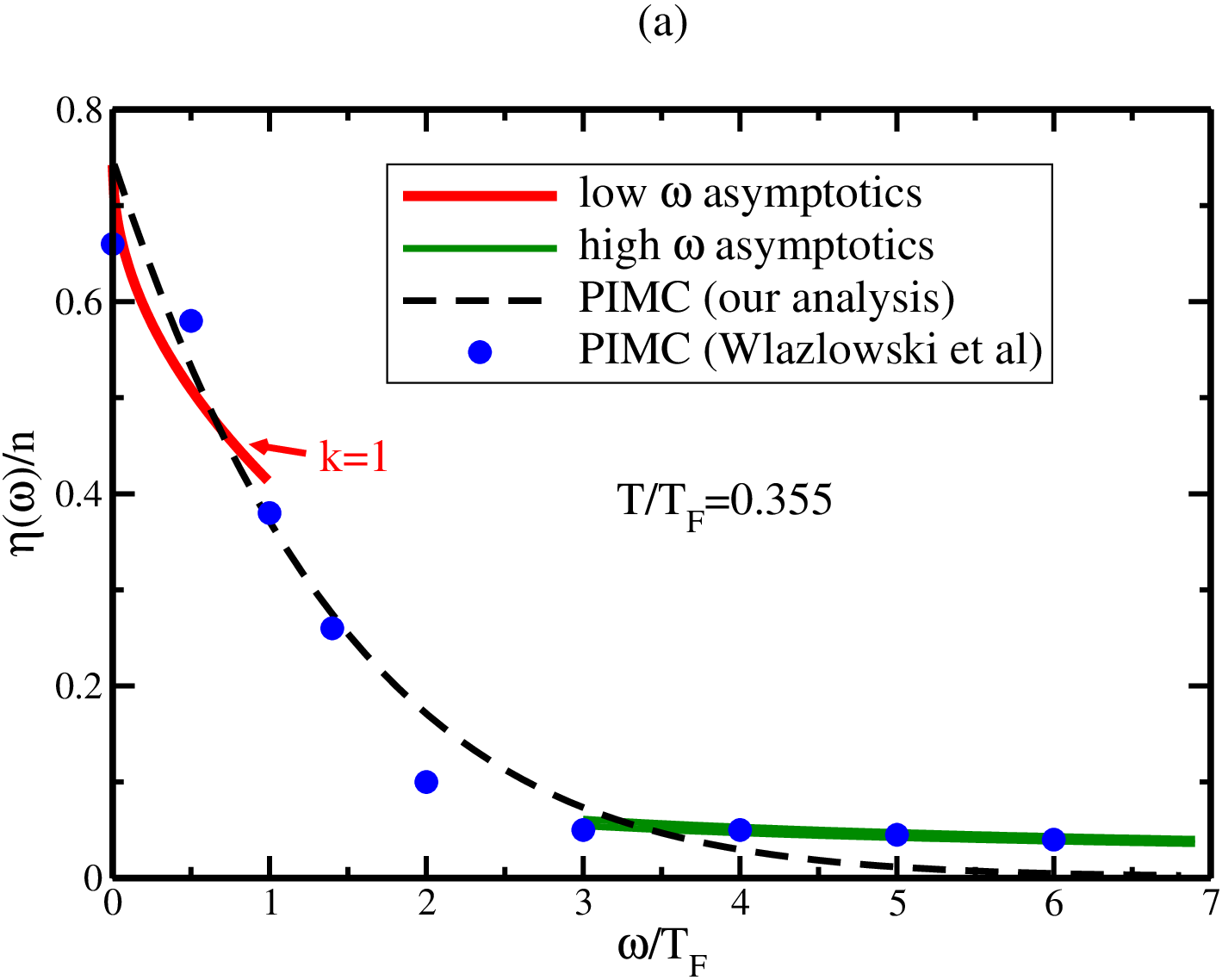}
\hfill
\includegraphics[width=0.45\linewidth]{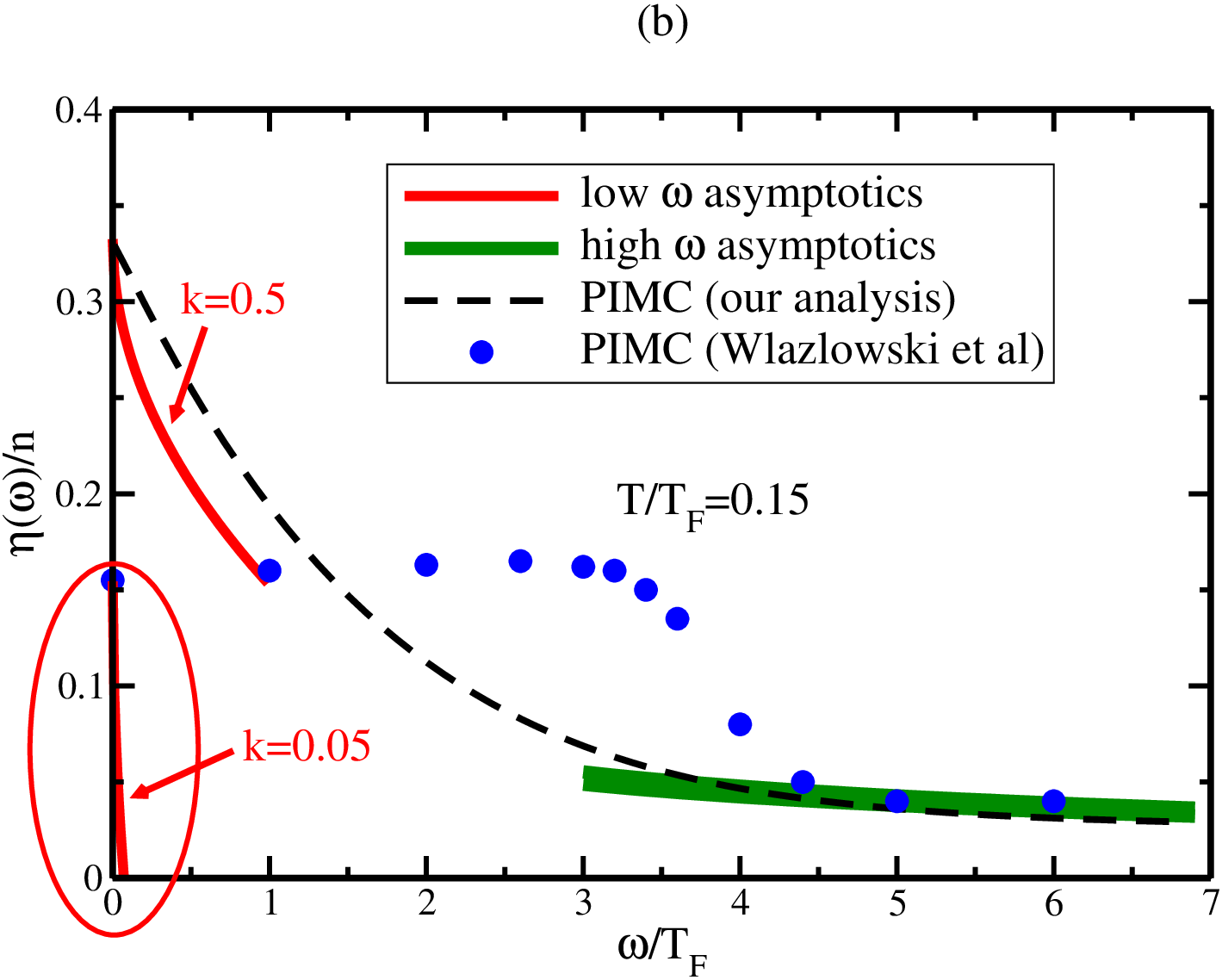}
\end{center}
\caption{Spectral function $\eta(\omega)/n$ extracted from PIMC simulation
Ref.~\cite{Wlazlowski:2012jb} with $N_x=8$, $N_\tau=52$ for $T/T_F=0.355$ (plot (a)) and $T/T_F=0.15$ (plot (b)). Dots: (approximate) result published in Ref.~\cite{Wlazlowski:2012jb}. Dashed: our re-analysis of the same data using a simple $\chi^2$ fit (Eq.~(\ref{mini})). See text for details.
Also shown is the known UV behavior (``high $\omega$ asymptotics''). Solid lines are low frequency behavior extracted from the known hydrodynamic behavior Eq.~(\ref{specfunc}), with values of $k$ indicated in the graphs. Since the $T/T_F=0.15$ low frequency peak is extremely narrow, we have highlighted it using an ellipse in plot (b). }
\label{fig:spectral}
\end{figure}

Let us perform this analysis on the spectral function for the shear viscosity
obtained with the Path Integral Monte Carlo (PIMC) approach from Ref.~\cite{Wlazlowski:2012jb}. In this article, the authors analyzed data using the Maximum Entropy Method (among others) based on the 5-parameter model class
$$
\eta(\hat \omega,m,\sigma,c,\alpha_1,\alpha_2)=f(\hat\omega,\alpha_1,\alpha_2)\frac{C}{15 \pi \sqrt{\hat\omega}}+\left(1-f(\hat\omega,\alpha_1,\alpha_2)\right) N(\hat\omega,m,\sigma,c)\,,
$$
where $f=e^{-\alpha_1 \alpha_2}\frac{e^{\alpha_1 \hat\omega}-1}{1+e^{\alpha_1 (\hat\omega-\alpha_s)}}$, $N=\frac{c}{\sqrt{2 \pi \sigma^2}} e^{-(\hat\omega-m)^2/2\sigma^2}$,
$\hat \omega=\omega/T_F$ and $C$ is Tan's contact (which is determined by other means and will be assumed to be known). This class of models implements the known large frequency behavior because $f(\omega\rightarrow \infty)\rightarrow 1$. Interestingly, note that the model also is compatible with the low frequency behavior in Eq.~(\ref{specfunc}), because $f(\hat\omega\ll 1)=\frac{\alpha_1}{1+e^{\alpha_1\alpha_2}} \hat\omega$. It is therefore possible that this model reproduces the correct low-frequency behavior, even though it is not guaranteed.
We re-analyze the data from Ref.~\cite{Wlazlowski:2012jb} for the correlator $G(\tau)$ for the $N_x=8,N_\tau=52$ simulation on $T/T_F=0.355$ using a simple $\chi^2$ minimization of
\begin{equation}
\label{mini}
G(\tau)-\frac{1}{\pi}\int_0^\infty d\hat\omega \eta(\hat\omega,m,\sigma,c,\alpha_1,\alpha_2)
\hat \omega \frac{\cosh(\hat\omega(\tau T_F-\frac{T_F}{2 T}))}{\sinh(\hat \omega T_F/2/T)}\rightarrow 0\,,
\end{equation}
neglecting the first three values of $\tau$ for reasons explained in Ref.~\cite{Wlazlowski:2012jb}. We find a very small $\chi^2/{\rm dof}\sim 4\times 10^{-3}$, but we also observe that the minimum is very broad in parameter space ($\chi^2$ does not change appreciably when changing parameters by order one). The extracted spectral function is shown in Fig.~\ref{fig:spectral}(a). As can be seen from this figure, in our re-analysis of the PIMC data, the extracted shape of $\eta(\omega)$ is close to the result reported in Ref.~\cite{Wlazlowski:2012jb}, but the tail asymptotics are only matched at much higher values of frequency. We find that the tail asymptotics are matched at $\omega\simeq 4 T_F$ when cutting the frequency integration at some pre-defined value of $\omega_{\rm cut}$, as done in Ref.~\cite{Wlazlowski:2012jb}.

Nevertheless, in both cases the low frequency behavior, including the extracted viscosity coefficient $\eta\equiv 
\lim_{\omega\rightarrow 0}\eta(\omega)$, is rather similar. Using the viscosity coefficient to extract a value for quantity $\eta_{\rm cl}/n$ as described above (assuming $k\simeq 1$), we can then compare the PIMC results to the purely hydrodynamic result Eq.~(\ref{specfunc})\footnote{In general, the extraction of $\eta_{\rm cl}/n$ will not be single-valued. While negative values of $\eta_{\rm cl}/n$ are clearly disallowed, we choose the smallest non-negative value of $\eta_{\rm cl}/n$.}. We find that for the case of $T/T_F=0.355$, the spectral function shape extracted from the PIMC data agrees reasonably well with the behavior demanded from Eq.~(\ref{specfunc}). Because it matches the known low frequency behavior, we conclude that this provides additional evidence that the extracted value of the shear viscosity of a unitary Fermi gas at $T/T_F=0.35$ is $\eta/n\simeq 0.7$.

Let us now repeat our analysis for the low temperature case $T/T_F=0.15$.
Simple minimization of Eq.~(\ref{mini}) leads to the result shown in
Fig.~\ref{fig:spectral}(b). Comparison to the result from 
Ref.~\cite{Wlazlowski:2012jb} now indicates that both methods of analysis agree with each other for large frequencies, while there is considerable discrepancy at low frequencies. While Ref.~\cite{Wlazlowski:2012jb} finds a spectral
function that is essentially featureless at low frequency, our re-analysis of the same data shows a peak at $\omega=0$. Trying to convert the viscosity coefficient $\eta(\omega=0)$ into a value for $\eta_{\rm cl}/n$, we find that, in both cases, the constant $k$ has to be less than unity. In particular, for our re-analysis we need $k\simeq 0.5$, while  $k\sim 0.05$ is needed to reach the viscosity value found in Ref.~\cite{Wlazlowski:2012jb}. The latter value is a much
larger deviation from unity than would be expected. Nevertheless, with these choices for $k$, we can evaluate the low frequency behavior of the spectral function, finding the results shown in Fig.~\ref{fig:spectral}(b). For our re-analysis, the low frequency shape is roughly consistent with the hydrodynamic behavior, although the agreement is somewhat less than satisfactory. 

For the results for the viscosity coefficient from Ref.~\cite{Wlazlowski:2012jb}, we find that the hydrodynamic behavior in Eq.~(\ref{specfunc}) results in an extremely narrow peak, which is in complete disagreement with the spectral function shape found in Ref.~\cite{Wlazlowski:2012jb}. We suspect that if indeed such a narrow peak is present in the physical spectral function, it could have been missed by the method used in  Ref.~\cite{Wlazlowski:2012jb} to reconstruct the spectral function from the imaginary time correlation function. If we simply add on such a low frequency peak to the broad spectral function found in Ref.~\cite{Wlazlowski:2012jb}, we find that the resulting value for the shear viscosity at $T/T_F=0.15$ is $\eta/n\simeq 0.35$, consistent with our re-analysis of the PIMC data, and consistent with the viscosity bound found in Ref.~\cite{Chafin:2012fu}.

\section{Discussion and Conclusion}
\label{sec:discussion}

\begin{figure}
\begin{center}
\includegraphics[width=0.45\linewidth]{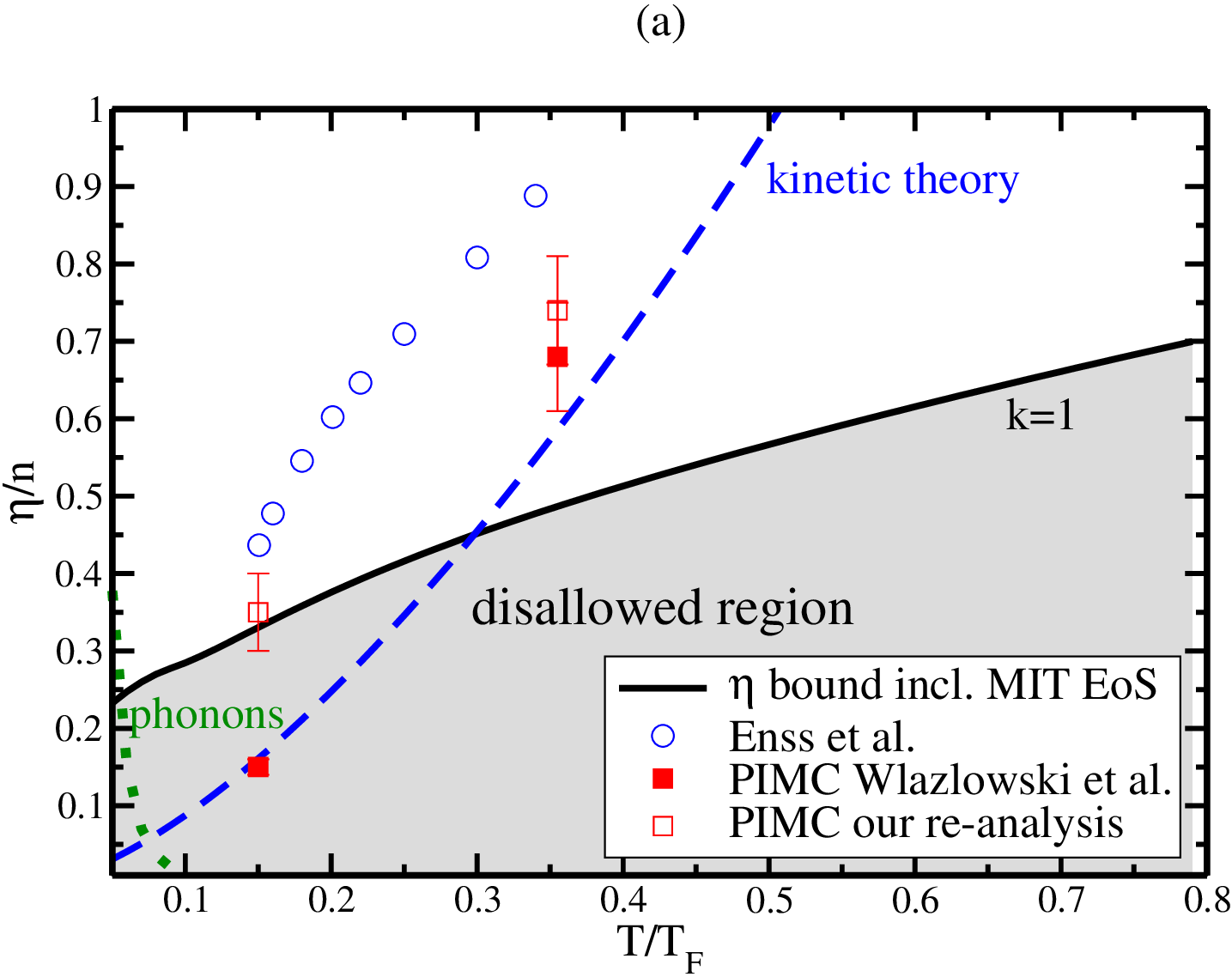}
\hfill
\includegraphics[width=0.45\linewidth]{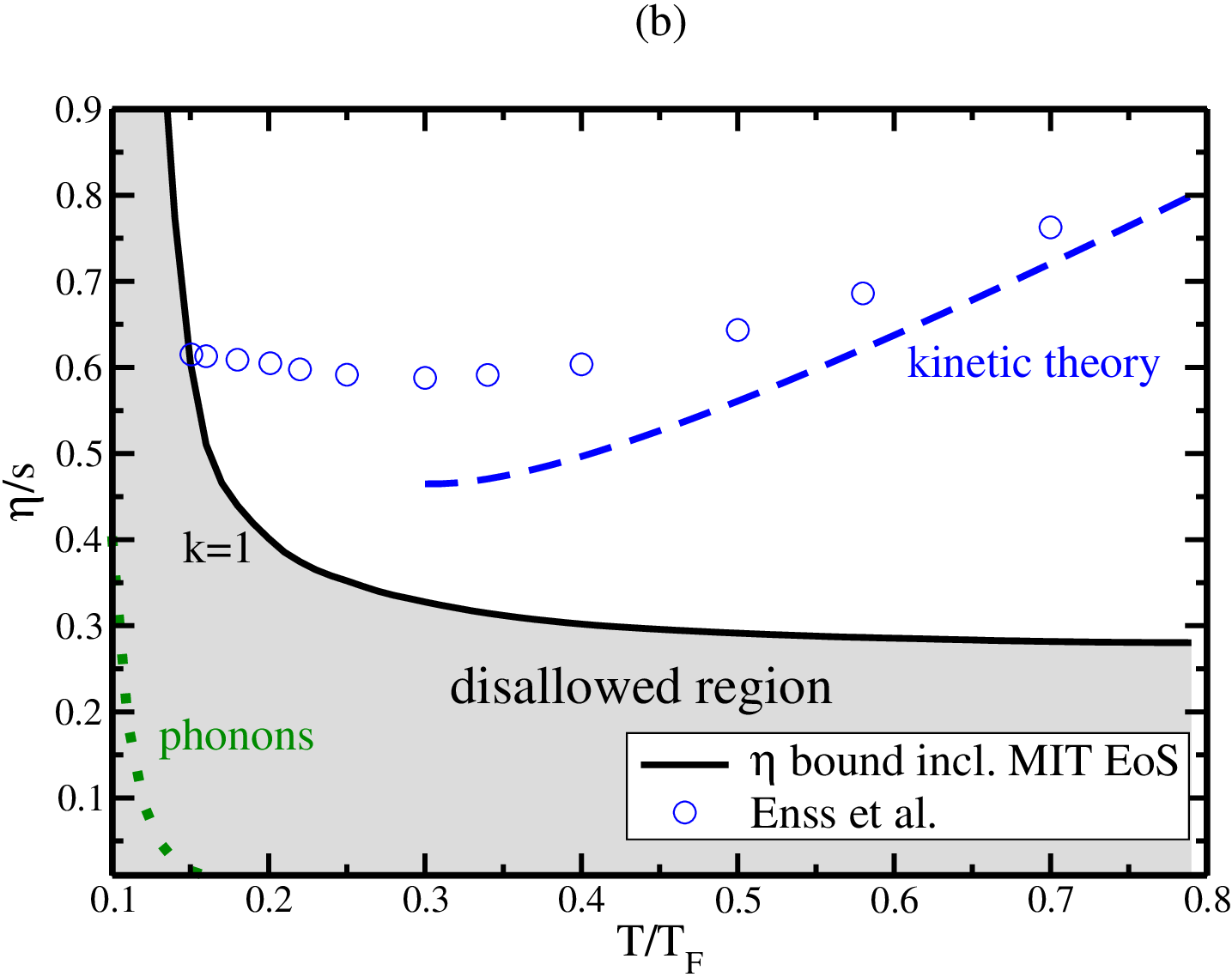}
\end{center}
\caption{Summary of results for $\eta/n$ (left) and $\eta/s$ (right). The viscosity limit from Eq.~(\ref{eq:viscmin3}) was used to delineate a ``disallowed region'' of viscosity values using the MIT equation of state from Ref.~\cite{Zwierlein} for $P/nT$ and $S/N$ and the estimate $k=1$. Also shown are results from kinetic theory at high temperatures and the phonon scattering at low temperatures (see text for details), as well as results from Enss et al., Ref.~\cite{Enss:2010qh}. Furthermore, we indicate the results from the PIMC calculation by Wlazlowski et al. \cite{Wlazlowski:2012jb} and our own re-analysis of the same data. Note that the resulting value for $\eta/n$ is consistent with the ``upper bound'' reported in the supplement to Ref.~\cite{Wlazlowski:2012jb}.}
\label{fig:minvisc}
\end{figure}

Employing the framework of hydrodynamics in the presence of fluctuations,
the low frequency behavior to the shear viscosity spectral function
has been calculated. The calculation gives a bound on the physical
shear viscosity, Eq.~(\ref{eq:viscmin3}), which matches the result found in Ref.~\cite{Chafin:2012fu} but disagrees with the low-temperature results from Ref.~\cite{Wlazlowski:2012jb}. We pointed out that the reason for the discrepancy can be tracked to the low-frequency shape of the spectral function extracted in Ref.~\cite{Wlazlowski:2012jb}. Specifically, assuming that the shape from Ref.~\cite{Wlazlowski:2012jb} is correct for $\omega>0.1 T_F$, hydrodynamic fluctuations predict a narrow but high peak at $\omega<0.1 T_F$, which effectively doubles the viscosity value. The situation is summarized in Fig.~\ref{fig:minvisc}, where the 
viscosity bound Eq. (\ref{eq:viscmin3}) is shown together with the viscosity in kinetic theory (cf.~Ref \cite{Massignan:2005zz})
\begin{equation}
\left.\frac{\eta}{n}\right|_{KT}=\frac{45 \pi^{3/2}}{32\times 2^{3/2}} 
\left(\frac{T}{T_F}\right)^{3/2}\simeq 2.768 \left(\frac{T}{T_F}\right)^{3/2}
\end{equation} 
as well as the viscosity of (inelastic) phonon scattering in the superfluid phase,
(cf.~Ref \cite{Mannarelli:2012eg,Braby:2010ec})
\begin{equation}
\left.\frac{\eta}{n}\right|_{phonons} = \left.\frac{\eta}{s}\right|_{phonons} \frac{s}{n}
\simeq 1.18\times 10^{-7} \left(\frac{T_F}{T}\right)^{5}\,.
\end{equation} 
We conclude that once the low frequency peak in the spectral function is included in the analysis, the viscosity value from the PIMC method is no longer in conflict with the lower viscosity limit Eq. (\ref{eq:viscmin3}). Also,
note that in this case the viscosity values are then much closer to the result from Enss et al. reported in Ref.~\cite{Enss:2010qh}.

Also shown in Fig.~\ref{fig:minvisc} are results for the minimum $\eta/s$,
using the experimental data from Ref.~\cite{Zwierlein} for the entropy per particle as a function of temperature. The corresponding results for kinetic theory and phonons are Ref.~\cite{Rupak:2007vp,Schafer:2009dj,Mannarelli:2012eg}
\begin{equation}
\left.\frac{\eta}{s}\right|_{KT}=2.768 \left(\frac{T}{T_F}\right)^{3/2} 
\frac{1}{\frac{5}{2}+\ln\left[\frac{3 \sqrt{\pi}}{4} (T/T_F)^{3/2}\right]}\,,\quad
\left.\frac{\eta}{s}\right|_{phonons}=4\times 10^{-9} \left(\frac{T_F}{T}\right)^{8}\,.
\end{equation} 
One observes that the viscosity result from kinetic theory enters the disallowed region at around $T/T_F\lesssim 0.3$. We interpret this as meaning that kinetic theory loses its validity at or above this temperature. Similarly, the viscosity result from the superfluid analysis enters the disallowed region for $T/T_F\gtrsim 0.1$. Again, we interpret this as meaning that the calculation loses validity at or below this temperature. The results for the shear viscosity by Enss et al. stay within the allowed region for all temperatures shown, but it is interesting to note that the $\eta/s$ values at low temperatures stop at the boundary to the disallowed region.

We conclude that the theory of hydrodynamic fluctuations can provide a relevant tool to study the properties of unitary Fermi gases. Specifically, we expect that our findings may help in the analysis of future quantum Monte Carlo calculations. We plan to further develop this theory in the future.

\acknowledgments

We would like to thank T. Sch\"afer for sharing an early version of Ref.~\cite{Chafin:2012fu} with us, as well as for fruitful discussion. Furthermore, we would like to thank M.~Zwierlein and M.~Ku for providing us with the experimental data from Ref.~\cite{Zwierlein}. Moreover, we would like to thank G.~Wlazlowski for providing us the PIMC data for the correlator from Ref.~\cite{Wlazlowski:2012jb} and pointing out the supplemental material included in Ref.~\cite{Wlazlowski:2012jb} and T.~Enss for providing us with the results from Ref.~\cite{Enss:2010qh}. Finally, we would like to thank V.~Guarie for a careful reading of this manuscript and valuable comments.

\appendix

\section{Finding the Velocity-Velocity Correlator}
\label{sec:finding}

Let us begin with the Navier-Stokes and continuity equations
\begin{equation}
\frac{\partial}{\partial t}(\rho v_k) +\frac{\partial T_{i k}}{\partial x_k}  =0\,,\quad
\frac{\partial }{\partial t} \rho +\frac{\partial }{\partial x_k} (\rho v_k)=0\,,
\end{equation}
where $T_{i k}$ is the standard viscous fluid stress tensor.  For consistency with conformal field theories the bulk viscosity, $\zeta $, has been set to zero.  A noise term, $ \Xi_{ij} $, is added to the stress tensor, $T_{ik} \rightarrow T_{ik}^{\rm cl}+\Xi_{ij}$ to account for short distance fluctuations.  Here we take $\Xi_{ij}$ to be given by short distance Gaussian noise:
\begin{eqnarray}
\langle \Xi_{i k} \Xi_{n m} \rangle &=&  \eta_{\rm cl} T 
\left(\delta_{i n} \delta_{k m}+\delta_{im} \delta_{k n}-\frac{2}{3} \delta_{ik}\delta_{nm}\right) \delta (x-x')\,.
\end{eqnarray}

Now consider a system in equilibrium, and allow for small perturbations away from equilibrium in $v_k$ and $ \rho \rightarrow \rho_o+ \delta \rho $.  The equation of fluid mechanics to first order in the perturbations are 
\begin{equation}
\frac{\partial}{\partial t}(\rho v_k) +\frac{\partial }{\partial x_k}(T_{i k}+\Xi_{ik})  =0\,,\quad
\partial_t \rho +\rho_0 \frac{\partial v^k}{\partial x^k}=0\,.
\end{equation}
The velocity can be decomposed  into longitudinal and transverse parts satisfying
\begin{equation}
\nabla \cdot v_t (r,t) =0\,, \quad
\nabla \times v_l (r,t)=0\,.
\end{equation}
Calculationally, this is easiest to implement if we restrict quantities to depend only on one spatial coordinate, say the $z$ coordinate, and time, e.g. $v_k=v_k(z,t)$. Then $v_z$ corresponds to the longitudinal component and $v_{x,y}$ correspond to the transverse components of the fluid velocity. Note that the results using this choice will match the results found if we had not made any restrictions on the coordinate dependence.

Using these restrictions, one finds the following system of fluid dynamic equations: The linear form of the continuity equation yields
\begin{equation}
\frac{\partial \rho}{\partial t} + \rho_0 \nabla \cdot v_l =0
\end{equation}
and momentum conservation yields the following system of equations
\begin{equation}
\frac{\partial }{\partial t} v_t - \frac{ \eta_{\rm cl}}{\rho_0} \Delta v_t=-\frac{1}{\rho_o}\partial_z \Xi_{x z}\,,\quad 
\rho_0 \frac{\partial }{\partial t} v_l = - \nabla \delta P + \frac{4}{3} \eta_{\rm cl} \nabla (\nabla \cdot v_l)-\partial_z \Xi_{zz}\,.
\end{equation}
The velocities can be solved for by taking a mixed Fourier-Laplace transform.  For simplicity, we will consider the transverse velocity:
\begin{equation}
v_t=\frac{1}{\rho_o}\frac{-i k \Xi_{x z}}{-i\omega +\gamma k^2}\,,
\end{equation}
where $ \gamma = \eta_{\rm cl} / \rho $. The symmetric correlator is
\begin{equation}
\langle v_i ^t v_j ^t \rangle =\frac{2}{\rho_o^2} (\delta_{i j} - \frac{k_i k_j}{k^2}) \frac{  k^2 \langle (\Xi_{x z})^2 \rangle } {\omega ^2 +\gamma^2 k^4}
\end{equation}
Evaluating the noise term, we find $\langle (\Xi_{x z})^2 \rangle=\eta_{\rm cl}T$ and hence
\begin{equation}
\label{corr-vt-vt}
\langle v_i ^t v_j ^t \rangle =\frac{2T}{\rho_o} (\delta_{i j} - \frac{k_i k_j}{k^2}) \frac{ \gamma k^2  } {\omega ^2 +\gamma^2 k^4}\,.
\end{equation}

Repeating this process for the longitudinal velocity, the entire velocity-velocity correlator is found to be
\begin{equation}
\label{corr-action5}
\Delta_{ij}^S \equiv \langle v_i v_j \rangle = \frac{2T}{\rho_o} \bigg[\frac{k_i k_j}{k^2} \frac{ \tilde{\gamma} \omega^2 k^2}{(\omega^2-c^2 k^2)^2+{(\tilde{\gamma} \omega^2 k^2)^2}} +\bigg(\delta_{i j} - \frac{k_i k_j}{k^2} \bigg) \frac{ \gamma k^2  } {\omega ^2 +\gamma^2 k^4} \bigg]\end{equation}
where $ \tilde{\gamma} = \frac{4}{3} \gamma $. Note that this is the symmetric correlator, as indicated by the superscript $S$. Now let us calculate the retarded velocity-velocity correlator. The classical fluctuation-dissipation theorem reads
\begin{equation}
\chi'' (\omega)=\frac{1}{2} \beta \omega C(\omega)
\end{equation}
with $\beta=1/T$ being the inverse temperature and
\begin{equation}
C_{i j}(t) = \langle A_i (t) A_j (0) \rangle _c = \langle  \delta A_i (t) \delta A_j (0) \rangle\,.
\end{equation}
This is the Kubo or relaxation function where the subscript $c$ stands for cumulant or connected.  In Fourier space $C_{ij}(\omega) $  is the symmetric correlator of the fluctuations. 
%

To find the retarded correlator, we must examine the generalized susceptibility. It is related to the time derivative of the Kubo function by
\begin{equation}
\chi_{i j} (t)= - \beta \Theta (t) \dot{C}_{i j} (t)
\end{equation}
where $ \Theta (t) $ is a step-function.  The generalized susceptibility $\chi$ can be identified as the retarded correlator. Utilizing the relation $\chi''=\text{Im}(\chi)$ and that $\chi$ is analytic in the upper half plane we can find the retarded velocity correlation function:
\begin{equation}
\label{eq:corr-retarded}
\Delta_{ij}^R=\frac{1}{\rho_o}\bigg[ \frac{k_i k_j}{k^2} \frac{ \omega^2 }{(\omega^2 -c^2 k^2)^2+(i \omega k^2 \tilde{\gamma})}+\bigg(\delta_{i j}-\frac{k_i k_j}{k^2} \bigg)\frac{ -\gamma k^2 }{-i \omega + \gamma k^2}\bigg]
\end{equation}

The  Kubo-Martin-Schwinger conditions relate the symmetric and retarded correlators in  quantum mechanical regimes. In Fermi systems, the response functions are related to bosonic operators (e.g. bilinears of fermi operators).  Therefore, a classical version of the KMS relation is found to be $G_s= \frac{2T}{\omega} \text{\rm Im}(G_R)$.  

\section{Equation of State}
\label{sec:eos}

\begin{figure}
\begin{center}
\includegraphics[width=0.45\linewidth]{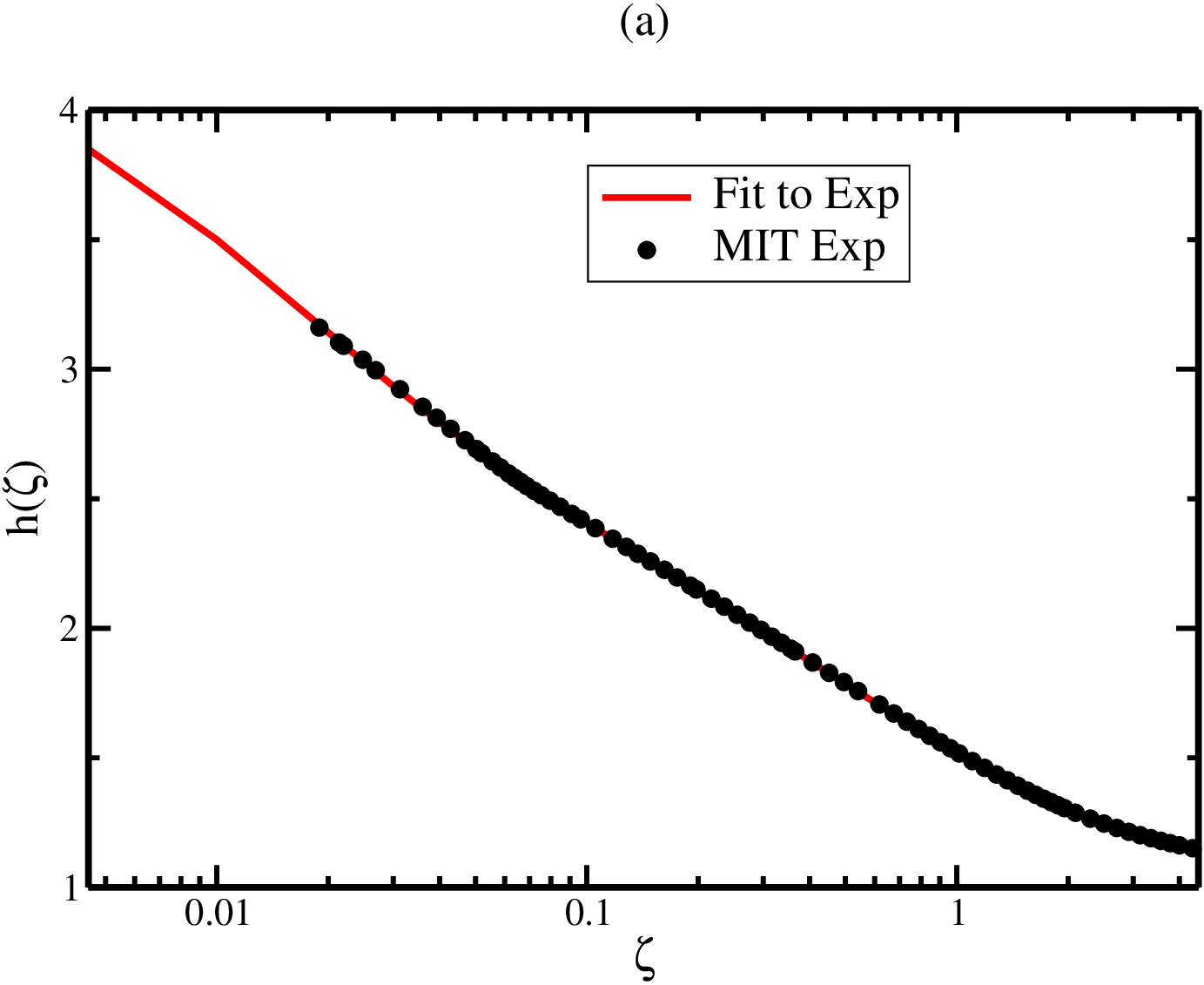}
\hfill
\includegraphics[width=0.45\linewidth]{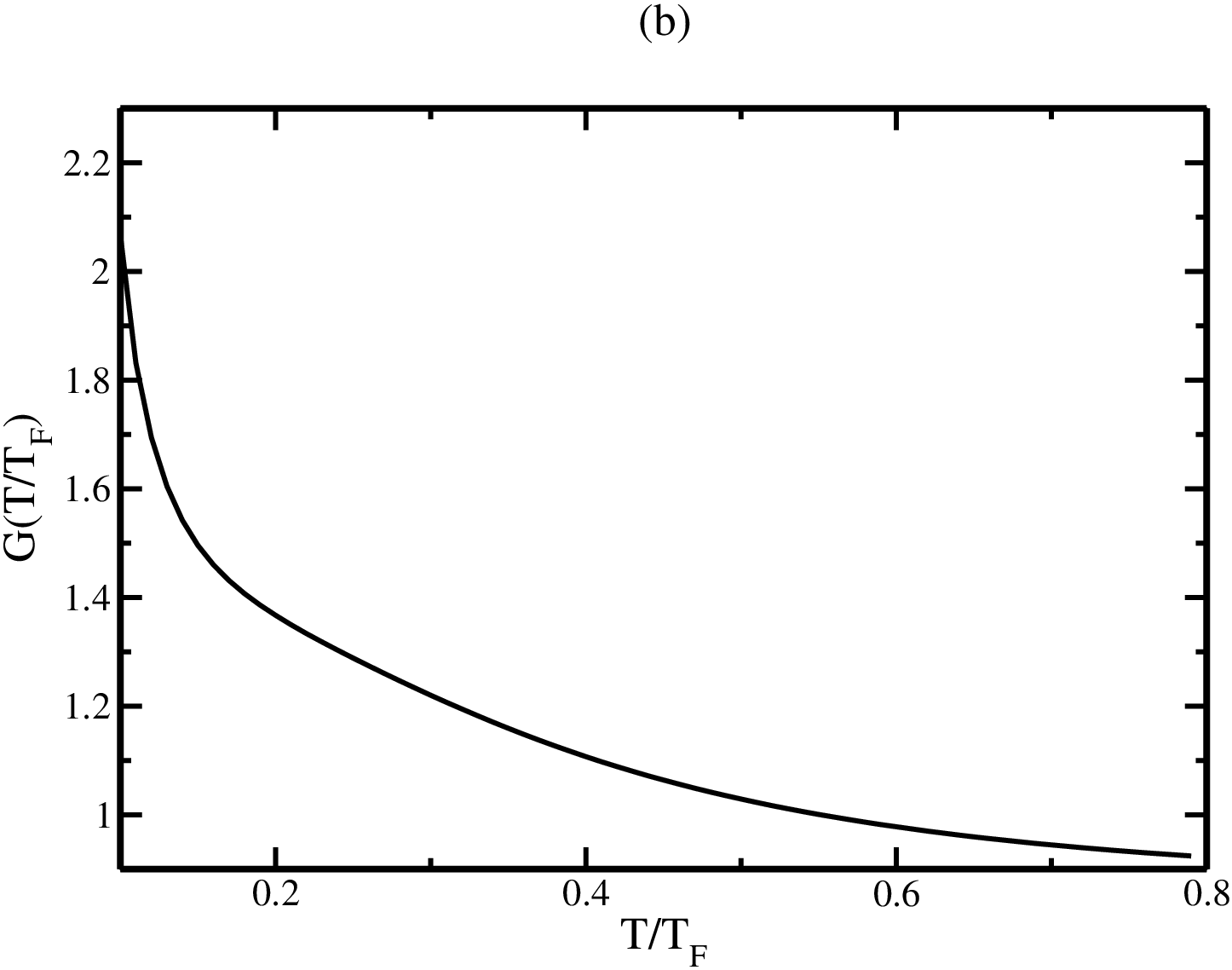}
\end{center}
\caption{Left: results for the normalized pressure $P/P_0$ from the MIT experiment Ref.~\cite{Zwierlein} and a 5 parameter fit (see text for details). Right: The function $G=\frac{P}{n T}$ as a function of $T/T_F$.}
\label{fig:EoS}
\end{figure}

The equation of state of a cold unitary Fermi gas was measured experimentally by the MIT group \cite{Zwierlein}. The result from Ref.~\cite{Zwierlein} for the pressure (versus fugacity $\zeta=e^{-\mu/T}$) is shown in Fig.~\ref{fig:EoS}(a). The result for the pressure is normalized by the ideal non-interacting single-component Fermi gas pressure,
$$
P_0(\mu,T)=-T \lambda_{dB}^{-3} {\rm Li}_{5/2}(-\zeta^{-1})\,,
$$
where $\lambda_{dB}=\sqrt{\frac{2 \pi}{m T}}$ is the deBroglie wavelength.
Also shown in Fig.~\ref{fig:EoS}(a) is a fit of the experimental data using the parametrization
$$
h(\zeta)\equiv P(\mu,T)/P_0(\mu,T)=\frac{\zeta^3+c_1 \zeta^2 +c_2 \zeta + c_3}{\zeta^3+c_4 \zeta^2 + c_5 \zeta + 1}
$$
which is a variant of the approach followed in Ref.~\cite{Schafer:2010dv}. A least squares fit gives
$$
c_1=101.591\,,\quad c_2=158.375\,,\quad c_3= 4.30128\,,\quad
c_4=104.884\,,\quad c_5=67.3894\,.
$$
Note that $c_3$ is related to the Bertsch parameter $\xi$ as 
$c_3=\xi^{-3/2}$ from the relation $P(\mu,T=0)=\xi^{-3/2} P_0(\mu,T=0)$.
The fitted value of $c_3$ therefore corresponds to $\xi=0.378$.
Using the thermodynamic relation $n=\left.\frac{\partial P}{\partial \mu}\right|_T$ and the identity $n=\lambda_{dB}^{-3} \frac{(4 \pi)^{3/2}}{3 \pi^2}\left(\frac{T_F}{T}\right)^{3/2}$, one has
$$
 \frac{(4 \pi)^{3/2}}{3 \pi^2}\left(\frac{T_F}{T}\right)^{3/2}=\zeta \frac{\partial}{\partial \zeta} \left({\rm Li}_{5/2}(-\zeta^{-1}) h(\zeta)\right)\,,
$$
and hence a (numerical) result for $\zeta=\zeta(T/T_F)$ by inverting the 
above functional relation.

Using the above relations, the pressure may be written as
$$
P(\mu,T)=n T G(\zeta)\,, \quad
G(\zeta)=\frac{1}{-\zeta \frac{d}{d\zeta} \ln\left[-{\rm Li}_{5/2}(-\zeta^{-1}) h(\zeta)\right]}\,.
$$
For convenience, the function $G$ is plotted in Fig.~\ref{fig:EoS}(b) as a function of $T/T_F$. As can be seen, the form of $G$ implies that the pressure is close to $n T$, but not monotonically larger or smaller than $n T$ in the domain considered. We have checked that $\lim_{T\rightarrow \infty} G(T/T_F)\rightarrow 1$, as expected.

From our parameterization, we can calculate thermodynamic quantities.  For example, the compressibility can be calculated by $\kappa=\frac{1}{n^2} \frac{\partial^2 P}{\partial \mu^2}|_T$.  As the compressibility is dependent on the second derivative of the pressure it contains a clear signature of the superfluid $\lambda$ phase transition Ref. \cite{Zwierlein}.  Another, thermodynamic quantity of interest is the entropy $s$.  This can be calculated from $s=\frac{\partial P}{\partial T}|_{\mu}$ or from $s/Nk_B=\frac{T_F}{T}(\frac{P}{P_0}-\frac{\mu}{E_F})$, where $E_F$ is the Fermi energy.


\end{document}